\documentclass{article}
\usepackage{spconf,amsmath,graphicx,hyperref,amsfonts, subcaption, xcolor, booktabs, multirow, cite}

\title{Rethinking Speech Representation Aggregation in Speech Enhancement: A Phonetic Mutual Information Perspective}

\name{Seungu Han$^{\star}$, Sungho Lee$^{\star}$, Kyogu Lee$^{\star \dagger}$}
\address{$^{\star}$ Department of Intelligence and Information, Seoul National University, Republic of Korea\\
         $^{\dagger}$ Artificial Intelligence Institute, Seoul National University, Republic of Korea}
\begin{document}
\ninept
\maketitle
\begin{abstract}
Recent speech enhancement (SE) models increasingly leverage self-supervised learning (SSL) representations for their rich semantic information. Typically, intermediate features are aggregated into a single representation via a lightweight adaptation module. However, most SSL models are not trained for noise robustness, which can lead to corrupted semantic representations. Moreover, the adaptation module is trained jointly with the SE model, potentially prioritizing acoustic details over semantic information, contradicting the original purpose.
To address this issue, we first analyze the behavior of SSL models on noisy speech from an information-theoretic perspective. Specifically, we measure the mutual information (MI) between the corrupted SSL representations and the corresponding phoneme labels,  focusing on preservation of linguistic contents.  
Building upon this analysis, we introduce the linguistic aggregation layer, which is pre-trained to maximize MI with phoneme labels (with optional dynamic aggregation) and then frozen during SE training. 
Experiments show that this decoupled approach improves Word Error Rate (WER) over jointly optimized baselines, demonstrating the benefit of explicitly aligning the adaptation module with linguistic contents.
\end{abstract}

\section{Introduction}
\label{sec:intro}
Recent speech enhancement (SE) methods \cite{wsum_eval, wsum_acoustic_se, finally, genhancer, ditse} have increasingly integrated self-supervised learning (SSL) representations of speech, motivated by their ability to capture high-level structure and thereby information of linguistic content and speaker characteristics. However, these SSL models are generally trained on clean data without considering robustness, and there is no guarantee that they retain their semantic information when the input signal is corrupted. 
While research exists on robust speech representations \cite{robustdata2vec, robustdistiller, dehubert}, fine-tuning these large-scale models on noisy data is often impractical. This is due to the high computational cost, the risk of degrading performance on clean speech, and reduced generalization capabilities, which is a primary goal of such representations.

Therefore, to quantify this degradation, we first analyze how linguistic information is distributed and preserved across the model's layers when the input is corrupted.
We employ the information-theoretic framework from \cite{mutual}, measuring the lower bound of mutual information (MI) between the SSL features obtained from noisy speech and their corresponding linguistic contents (phonemes). 
We analyze three pre-trained Base models: HuBERT \cite{hubert}, WavLM \cite{wavlm}, and wav2vec 2 \cite{wav2vec2}.
Note that, while recent work studied the layer-wise distribution of linguistic properties for clean speech \cite{layerwise_w2v, layerwise_all}, an analysis of these representations under noisy conditions is needed to understand their practical robustness.
We find that, while the upper layers remain the primary carriers of phonetic information even in noisy conditions, the distribution varies with SNR and the absolute magnitude of peak information is reduced.

In SE, SSL representations are commonly used as an auxiliary input along with the noisy speech. A typical approach is to introduce a lightweight aggregation module, e.g., a learnable weighted sum of intermediate features \cite{wsum_eval, wsum_acoustic_se, genhancer, ditse}, which is then used to condition the SE model. Such a method was motivated by observations that different layers tend to capture distinct acoustic and linguistic properties \cite{layerwise_w2v, layerwise_all}. 
We observe two limitations of this approach.
First, the aggregation modules are typically trained jointly with the SE model using only ``acoustic'' criteria---ones that compare the SE prediction with clean ground truth in the low-level signal representations, such as magnitude spectrograms. As a result, the joint SE training could lack guidance to preserve or recover the rich semantic information from the SSL features. 
Indeed, our empirical analysis shows that the resulting aggregation weights from joint SE training differ from those that maximize the MI with the phonemes.
Note that, while the SSL model can alternatively be used for a semantic loss function \cite{ssl_loss, ssl_for_se, ssl_loss_for_ssl}, this method introduces significant computational overhead from an extra forward and backward pass through the SSL model during training.
Second, our investigation shows that the layer-wise information distribution varies with Signal-to-Noise Ratio (SNR), and so do the optimal adaptation weights. Since SNR is a time-varying characteristic, this suggests that the layer aggregation should also be dynamic. 

Building on this insight, we propose the Linguistic Aggregation Layer, a module specialized in preserving linguistic content. 
In our decoupled framework, this module is first pre-trained by maximizing the MI between its output and phoneme labels. Once linguistically optimized, the module is frozen and used to provide robust linguistic conditioning for the main SE model. We first explore this method with a global Weighted-Sum (WS) and find it effective at preserving linguistic information. 
Additionally, we propose a more powerful, time-varying Dynamic Weighted-Sum (DWS). Experiments show that this linguistic-first approach significantly reduces the Word Error Rate (WER) compared to jointly optimized baselines, demonstrating the benefit of explicitly aligning the conditioning module with its intended semantic purpose.

\begin{figure*}[t]
    \centering
    
    \begin{subfigure}{.33\linewidth}
        \centering
        \includegraphics[width=\linewidth]{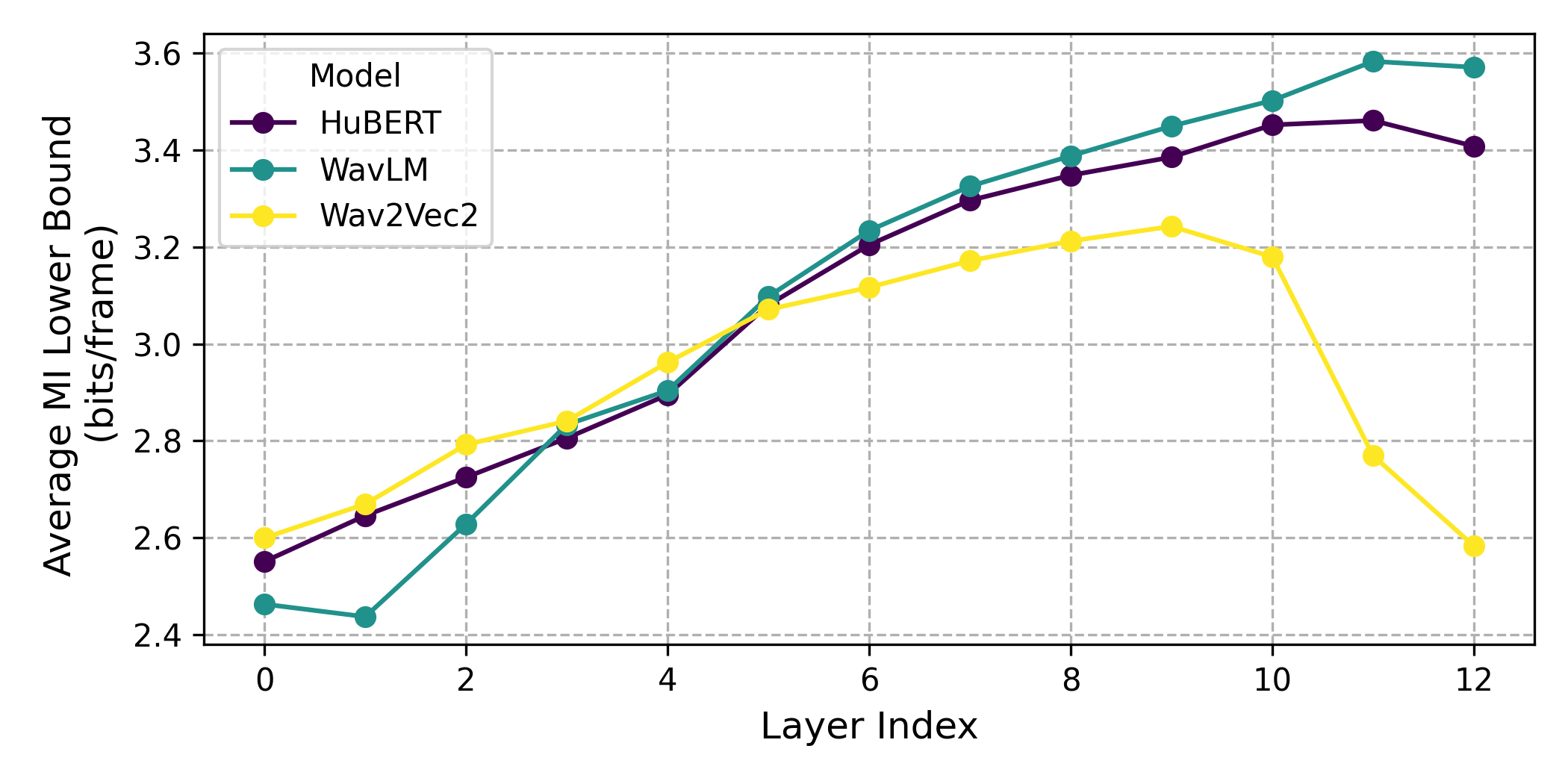}
        \subcaption{Layer-wise MI for various models.}
        \label{fig:mi_per_layer_phoneme}
    \end{subfigure}
    \begin{subfigure}{.33\linewidth}
        \centering
        \includegraphics[width=\linewidth]{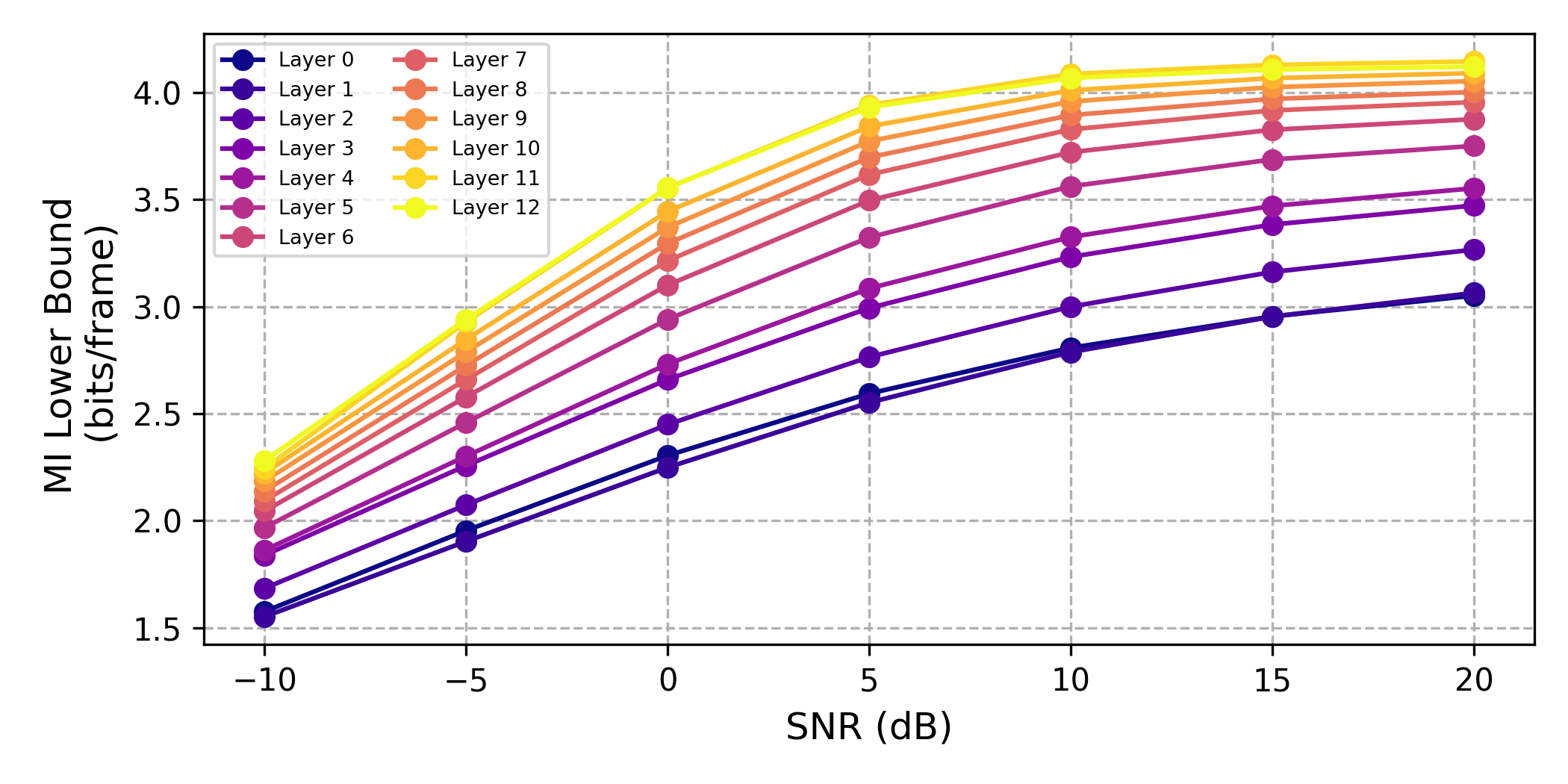}
        \subcaption{MI vs. SNR for each layer (WavLM).}
        \label{fig:mi_vs_snr_phoneme}
    \end{subfigure}
     \begin{subfigure}{.33\linewidth}
        \centering
        \includegraphics[width=\linewidth]{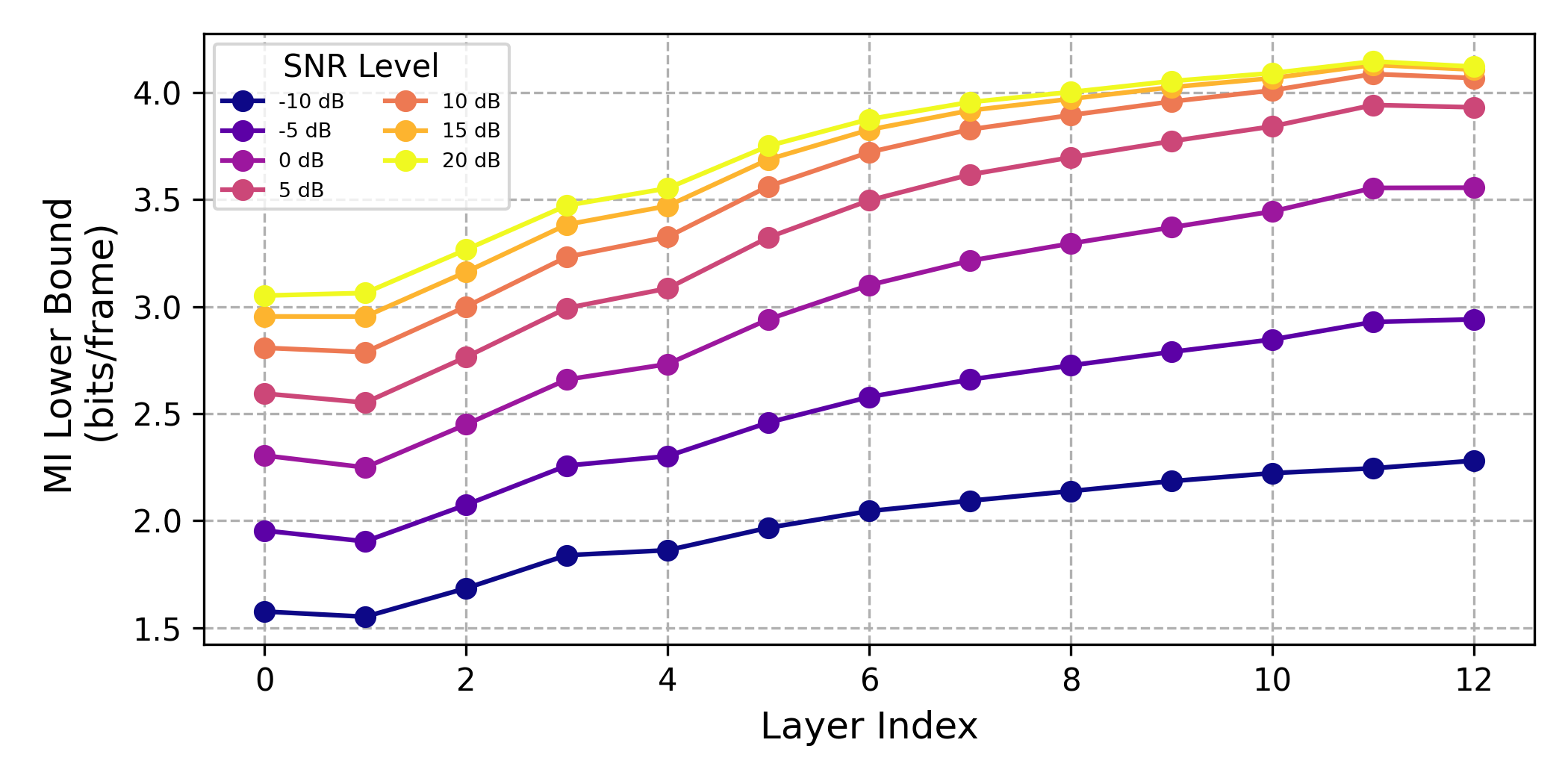}
        \subcaption{MI vs. Layer index for each SNR (WavLM).}
        \label{fig:mi_vs_layer_phoneme}
    \end{subfigure}\\
    
    \caption{
        Analysis of the MI lower bounds between the SSL features and phonemes across different layers and SNR conditions.
        In (a), "Average MI Lower Bound" denotes values averaged across all evaluated SNR levels.
    }
    \label{fig:mi_phoneme_main}
\end{figure*}

\section{Measuring Self-supervised Models' Robustness with Mutual Information}
\label{sec:method}
\looseness=-1
Liu et al. \cite{mutual} adopted an information-theoretic framework and proposed a practical method to estimate an SSL model's performance on downstream tasks.
We extend the application of this framework to a more challenging scenario, quantifying the noise robustness of SSL representations and their applications for the SE task.

\subsection{Bounding Mutual Information with Labeled Data}
\label{subsection:mi_eqs}
Most conventional MI estimation methods for speech representations \cite{clustering, speech_mi, hubert, layerwise_w2v} rely on clustering of the representations to construct empirical discrete probability distributions, due to the intractability of direct MI estimation for high-dimensional, continuous data. In our case, however, the presence of noise could further perturb the representations, preventing us from ensuring that the clusters are meaningful and effective for the MI quantification.

Therefore, we adopt the approach of \cite{mutual}, which instead lower-bounds MI through direct probing. 
Specifically, we consider an MI $I(Z;Y)$ between a noisy representation $Z$ and a target categorical label $Y$ (e.g., the corresponding phoneme). While the entropy of the target $H(Y)$ can be estimated from an empirical distribution, the full MI metric is intractable because the true conditional probability $p(y|z)$ is unknown.
To overcome this, \cite{mutual} introduce a linear probe, $q_\phi(y|z)$, to approximate $p(y|z)$. This allows us to derive a tractable lower bound on the MI as follows, 
\begin{subequations}
\begin{align}
I(Z;Y)
&= H(Y) - H(Y\mid Z) \\
&= H(Y) - \mathbb{E}_{(y,z)\sim p(y,z)}\bigl[-\log p(y\mid z)\bigr] \\
&= H(Y) - \mathbb{E}_{p}\left[
    -\log q_\phi(y|z)
    - \log\frac{p(y|z)}{q_\phi(y|z)}
  \right] \\
&\geq H(Y) - \mathbb{E}_{p}\left[-\log q_\phi(y|z)\right].
\end{align}
\end{subequations}
The last inequality follows from the non-negativity of the Kullback-Leibler divergence.
The final term, $\mathbb{E}_{p}\left[-\log q_\phi(y\mid z)\right]$, is the cross-entropy loss of the probe network. Therefore, by simply training the probe for the classification, we obtain the desired MI lower bound.

\subsection{Empirical Analysis}
\label{subsection:mi_setting}
We analyze three pre-trained SSL models with ``Base'' size: HuBERT \cite{hubert}, WavLM \cite{wavlm}, and wav2vec 2 \cite{wav2vec2}.
The probe network $q_\phi$ is a 3-layer multi-layer perceptron (MLP) with ReLU activation and dropout. Following \cite{mutual}, this probe is trained to predict force-aligned phoneme sequences \cite{forced_aligner} from noisy representations. These noisy representations are created by adding noise from folds 1-4 of the ESC-50 dataset \cite{esc50} to the clean audio from the dev-clean subset of LibriSpeech \cite{librispeech} with SNRs (-10, -5, 0, 5, 10, 15, and 20 dB). A trained MLP is used to estimate MI noisy test samples created using test-clean and a disjoint noise set from fold 5 of ESC-50. We optimize with a learning rate of 0.001 for 15 epochs.

Figure \ref{fig:mi_phoneme_main} shows the MI analysis results. Consistent with findings on clean speech \cite{layerwise_w2v, layerwise_all}, we observe that phonetic information peaks in the upper layers (9-11). 
Notably, wav2vec 2.0 exhibits an "autoencoder-style" behavior where, after deviating from input features, the deepest layers trend back toward the input \cite{layerwise_w2v}.
As expected, Figure \ref{fig:mi_vs_snr_phoneme} confirms that information degrades across all layers as noise increases, though the upper layers consistently retain more linguistic content. 
Figure \ref{fig:mi_vs_layer_phoneme} shows that the absolute magnitude of the peak is drastically reduced, suggesting that relying on a single layer may not provide sufficient linguistic information.

\section{Decoupled Linguistic Optimization for Speech Enhancement}
Conventional adaptation modules in SE, such as weighted sums of layer outputs, are typically trained jointly with the SE model using only universal objectives, such as simple spectrogram or waveform losses. While these objectives guide SE models to improve acoustic performance, they also guide the adaptation module to prioritize acoustically relevant features, neglecting its potential to preserve the semantic content encoded in the SSL representations. This joint training approach is standard practice, as in the SUPERB SE benchmark \cite{superb-sg, wsum_eval} and in subsequent work that improved acoustic performance by adding more acoustic features \cite{wsum_acoustic_se}.

To address this, we propose a decoupled approach that specializes the adaptation module for its linguistic role. The adaptation module is first trained independently to maximize the Mutual Information (MI) between its output and phoneme labels. This ensures that the module is specialized for retaining phonetic information. Once optimized, this linguistic module is frozen and used to condition the downstream SE model, which is then trained solely on acoustic objectives. This design allows the SE model to leverage both rich linguistic input. We focus on the widely-used weighted-sum architecture, which fuses the $L$ layer representations $\{L_i\}$ into a single representation $R_{\mathrm{fused}}$ using a set of scalar weights $\mathbf{w}$:
\begin{equation}
    R_{\mathrm{fused}}(\mathbf{w}) = \sum_i w_i L_i.
    \label{eq:wsum}
\end{equation}
The key difference between the conventional method and our proposal lies in how the weights $\mathbf{w}$ in Eq. (\ref{eq:wsum}) are determined.

\textbf{Acoustic-Tuned Weighted-Sum (WS):} This is the conventional approach where a weighted-sum adaptation module is jointly trained with the SE model using a standard acoustic loss. The layer weights are learned jointly with the SE model to minimize acoustic error, not to preserve semantics.

\textbf{Linguistic-Tuned Weighted-Sum (WS):} Our primary method determines the weights $\mathbf{w}$ on each layer differently. A set of fixed weights $\mathbf{w}^*$ is pre-trained by maximizing the MI between the fused representation $R_{\mathrm{fused}}(\mathbf{w})$ and the ground-truth phoneme labels $Y_{\mathrm{phoneme}}$. This process identifies which layers are consistently the most critical for linguistic content. Once learned, these weights are frozen during SE training.

\textbf{Hybrid Weighted-Sum (WS):} Building on the above method, this approach uses the same fixed linguistic weights $\mathbf{w}^*$ to combine the transformer layers, but allows the aggregation weight $w_0$ on the convolutional front-end of the SSL model to be fine-tuned with the SE model. This tests whether adapting the initial feature extraction helps while keeping the more linguistic layer fusion fixed.

\textbf{Dynamic Weighted-Sum (DWS):} Instead of assigning fixed scalar weights, we introduce a self-attention mechanism to capture time-dependent importance across layers. All hidden states from the $L$ layers are stacked at each time step, forming a tensor of shape $L\times D$ where $D$ is the feature dimension. A single-head self-attention is then applied across the layer dimension:
\begin{equation}
    R_{\mathrm{attn}} = \mathrm{Avg}_{\text{layer}}\!\left(\mathrm{Softmax}\!\left(\frac{QK^\top}{\sqrt{D}} + \mathbf{b}\right)V\right),
\end{equation}
where $Q,K$ are linear projections of the stacked layer representations, and $V$ is the stack itself. The $QK^\top$ term computes a dynamic $L\times L$ attention score matrix for each time frame. $\mathbf{b}$ is a time-invariant learnable bias vector of shape $1\times L$, encoding the layers' global relative importance. The attention output is then averaged across the layers to produce the final fused representation. This design allows the model to dynamically select informative layers at each time step while maintaining a global prior. Analogous to the weighted sum case, we also compare two training strategies: an acoustic-tuned DWS trained jointly with the SE model, and our linguistic-tuned DWS, which is first pre-trained by maximizing the lower bound of MI $I(R_{\mathrm{attn}}; Y)$ and frozen during SE training. For hybrid tuning, we make only bias of layer 0, $b_0$, learnable.

\label{sec:exp}
\subsection{Experimental Setup}
We follow the SE downstream task in the official SUPERB benchmark recipe\footnote{github.com/s3prl/s3prl/tree/main/s3prl/downstream/enhancement\_stft2} \cite{superb, superb-sg}, which uses the VoiceBank-DEMAND dataset \cite{vbd}. We utilize the same SE model as our backbone and replace only the standard adaptation module with our proposed, linguistically optimized versions. To investigate the impact of additional acoustic features, we also conduct experiments where a log1p spectrogram is concatenated with the SSL representation, following \cite{wsum_acoustic_se}. To obtain MI-maximizing aggregation modules, we jointly train them with linear probes using the same objective described in Section~\ref{subsection:mi_eqs}. We train the static linguistic WS module on the same dataset as in Section~\ref{subsection:mi_setting}. Regarding its higher complexity, the linguistic DWS module is trained on the larger train-clean-100 subset, augmented with MUSAN noise, to prevent overfitting.
We evaluate performance using two sets of metrics. For acoustic quality, we report standard metrics: SI-SDR \cite{sisdr}, STOI \cite{estoi}, and PESQ \cite{pesq}. To evaluate linguistic fidelity, we report the Word Error Rate (WER) obtained by transcribing the enhanced speech with a pre-trained Whisper-Large-v3\footnote{github.com/openai/whisper}. 
Note that WER can also be slightly affected by acoustic artifacts.

\subsection{Results}
\begin{figure}[t]
    \centering
    
    \begin{subfigure}[b]{\linewidth}
        \centering
        \includegraphics[width=\linewidth]{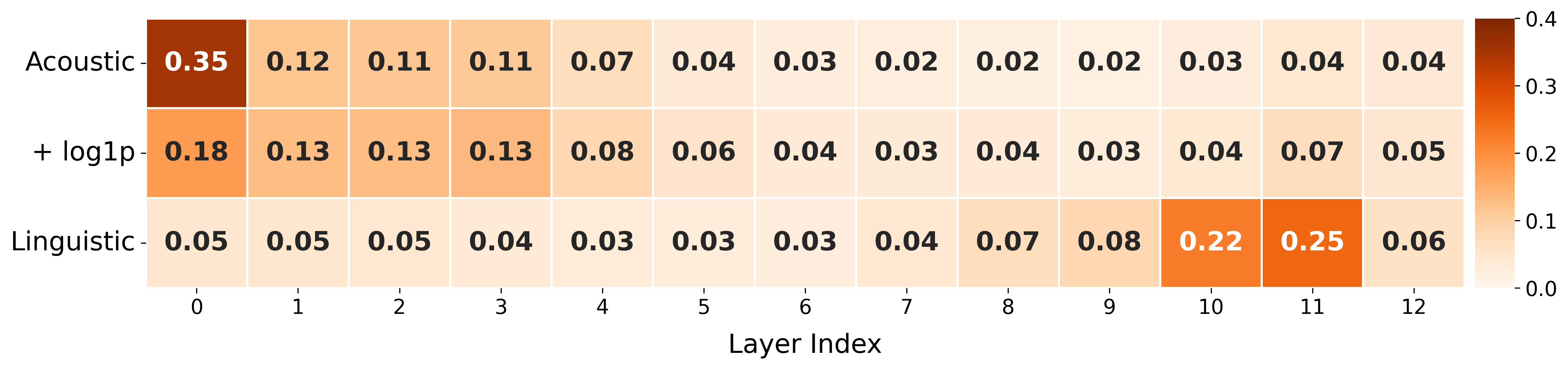}
        \subcaption{HuBERT}
        \label{fig:weights_hubert_col}
    \end{subfigure}
    

    \begin{subfigure}[b]{\linewidth}
        \centering
        \includegraphics[width=\linewidth]{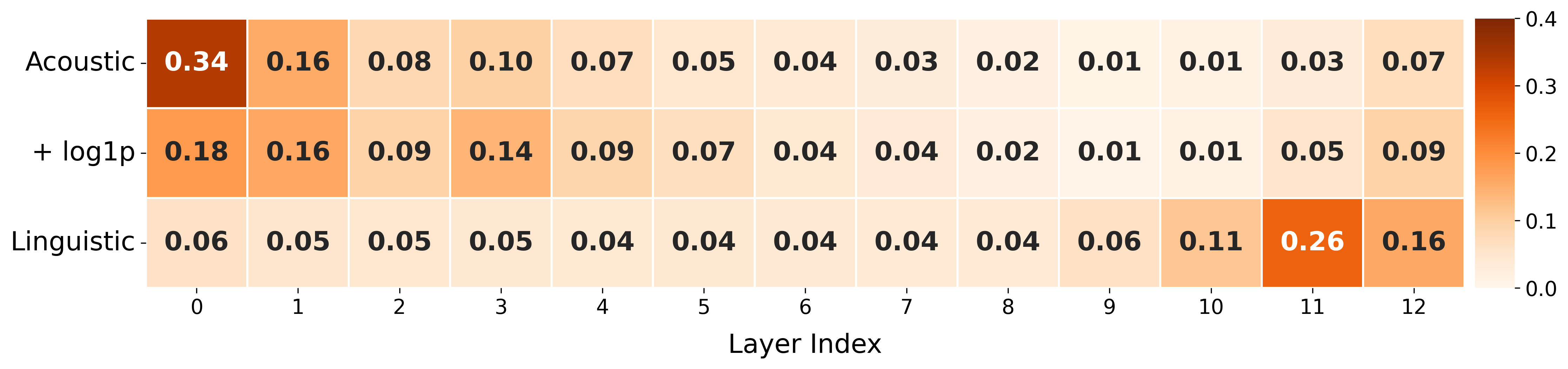}
        \subcaption{WavLM}
        \label{fig:weights_wavlm_col}
    \end{subfigure}


    \begin{subfigure}[b]{\linewidth}
        \centering
        \includegraphics[width=\linewidth]{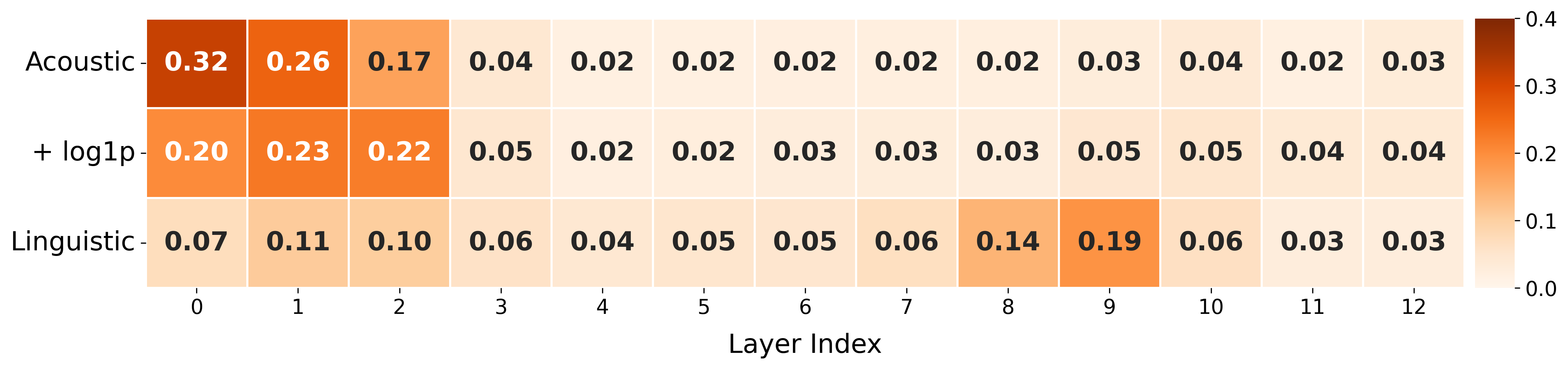}
        \subcaption{wav2vec 2}
        \label{fig:weights_wav2vec2_col}
    \end{subfigure}

    \caption{Layer aggregation weights used for each model.}
    \label{fig:all_weights_analysis_col}
\end{figure}

\begin{table}[t]
\centering
\caption{SE performance across pre-trained models (HuBERT, WavLM, wav2vec 2) using Weighted-Sum (WS) adaptation module.}
\label{tab:results_all_models}
\scalebox{0.75}{%
\begin{tabular}{l|l|c|ccc|c}
\toprule
\textbf{Model} & \textbf{Method} & \textbf{MI} $\uparrow$ &  \textbf{SI-SDR} $\uparrow$ & \textbf{STOI} $\uparrow$ & \textbf{PESQ} $\uparrow$ & \textbf{WER} $\downarrow$ \\
\midrule
\textbf{log1p} & & 1.05 & 9.47 & 0.945 & 2.81 & 0.0649 \\
\midrule
\multirow{7}{*}{\textbf{HuBERT}} & layer 11 & 3.47 & 8.99 & 0.945 & 2.82 & 0.0710 \\
& WS (Acoustic) & 3.30 & 9.63 & 0.949 & 2.97 & 0.0663 \\
& + log1p & - & \textbf{9.68} & \bf 0.951 & \textbf{3.06} & 0.0616 \\
& WS (Linguistic) & \textbf{3.49} & 8.85 & 0.947 & 2.93 & 0.0609 \\
& + log1p & - & 9.46 & 0.950 & 2.97 & 0.0648 \\
& WS (Hybrid) & 3.43 & 9.34 & 0.949 & 3.00 & 0.0590 \\
& + log1p & - & 9.26 & \bf 0.951 & 2.97 & \textbf{0.0586} \\
\midrule
\multirow{7}{*}{\textbf{WavLM}} & layer 11 & 3.57 & 8.90 & 0.946 & 2.84 & 0.0651 \\
& WS (Acoustic) & 3.47 & {9.44} & {0.949} & {2.99} & 0.0602 \\
& + log1p & - & 9.53 & 0.949 & {3.05} & {0.0585} \\
& WS (Linguistic) & \textbf{3.61} & 8.97 & 0.947 & 2.97 & 0.0596 \\
& + log1p & - & \bf 9.98 & \bf 0.950 & {3.04} & 0.0583 \\
& WS (Hybrid) & 3.57 & 9.37 & 0.948 & 2.95 & 0.0583 \\
& + log1p & - & 9.61 & \bf 0.950 & \bf 3.06 & \textbf{0.0517} \\
\midrule
\multirow{7}{*}{\textbf{wav2vec 2}} & layer 9 & 3.23 & 8.83 & 0.945 & 2.85 & 0.0665 \\
& WS (Acoustic) & 3.08 & 9.09 & 0.949 & 2.92 & 0.0625 \\
& + log1p & - & 9.42 & 0.949 & 2.99 & 0.0619 \\
& WS (Linguistic) & \textbf{3.32} & 8.98 & 0.948 & 2.92 & 0.0649 \\
& + log1p & - & 9.41 & \textbf{0.950} & \textbf{3.00} & 0.0630 \\
& WS (Hybrid) & 3.25 & \textbf{9.53} & 0.947 & 2.97 & \textbf{0.0603} \\
& + log1p & - & 9.47 & 0.947 & 2.96 & 0.0615 \\
\bottomrule
\end{tabular}%
}
\end{table}

\subsubsection{MI-optimized Weights Analysis}
Our MI analysis provides the direct motivation for our decoupled framework. The findings from Figure \ref{fig:mi_phoneme_main}, that phonetic information consistently peaks in the upper layers, suggest that an effective adaptation module must prioritize these layers. The layer aggregation weights in the weighted-sum module (Figure \ref{fig:all_weights_analysis_col}) confirm this principle. The baseline, when optimized for acoustic reconstruction, learns to heavily weight the acoustically-rich Layer 0 (after the convolutional frontend and before the main transformer layers). While incorporating the log1p spectrogram causes this focus to shift slightly towards the middle layers, potentially due to the log1p partially providing necessary acoustic information, it still largely prioritizes the use of the bottom half of the model. In contrast, our fixed Linguistic WS method, optimized for MI, correctly and consistently identifies and prioritizes the linguistically-rich upper layers.

\subsubsection{Speech Enhancement Results}
The performance results for the static WS module are shown in Table \ref{tab:results_all_models}. First, focusing on the methods using only the SSL representation, the WS (Acoustic) baseline consistently achieves the better performance on intrusive metrics, while our WS (Linguistic) method consistently yields a higher MI score and a significantly better WER except for wav2vec 2, although this comes at the cost of slight degradation on acoustic performance. This confirms our hypothesis that the linguistic information within SSL models can be effectively preserved and utilized simply by optimizing how the existing layers are aggregated. 
For the distinct behavior of the wav2vec 2 model, the MI analysis (Figure \ref{fig:mi_phoneme_main}) reveals that its phonetic information is not only lower overall but also lacks a distinct peak. Consequently, when we train the linguistic aggregation module, it learns a more distributed set of weights (Figure \ref{fig:weights_wav2vec2_col}). This lack of a clear linguistic peak within the representation provides a direct explanation for its relatively poor WER performance when used without additional acoustic features.

Next, we analyze the effect of incorporating the log1p spectrogram feature. While the acoustically-tuned baseline benefits from this additional input, naively concatenating the raw spectrogram with our specialized linguistic representation leads to inconsistent and often worse WER performance (Table \ref{tab:results_all_models}). We hypothesize this is because the downstream SE model struggles to effectively fuse features from two different levels of characteristics. To relieve this, we propose WS (Hybrid), which lets the model also use acoustic information of the SSL feature, by fine-tuning the SSL model's convolutional front-end (Layer 0) while keeping the linguistic fusion of the upper layers fixed. Its consistently strong WER suggests that the failure is not in the linguistic module itself, but in the naive fusion strategy, and that a linguistic-first approach is still effective.

\begin{table}[t]
\centering
\caption{SE performance using the Dynamic Weighted-Sum (DWS) adaptation module on WavLM. 
}
\label{tab:results_self-attn}
\scalebox{0.75}{%
\begin{tabular}{l|c|ccc|c}
\toprule
\textbf{Method} & \textbf{MI} $\uparrow$ &  \textbf{SI-SDR} $\uparrow$ & \textbf{STOI} $\uparrow$ & \textbf{PESQ} $\uparrow$ & \textbf{WER} $\downarrow$ \\
\midrule
DWS (Acoustic) & 3.23 & 9.42 & 0.948 & 2.98 & 0.0587 \\
+ log1p & - & 9.53 & \textbf{0.949} & 2.99 & \bf 0.0545 \\
DWS (Linguistic) & \textbf{3.68} & 9.09 & 0.946 & 2.89 & 0.0578 \\
+ log1p & - & 9.54 & \textbf{0.949} & 3.02 & 0.0584 \\
DWS (Hybrid) & 3.62 & \bf 9.63 & 0.948 & 2.93 & 0.0593 \\
+log1p & - & 9.61 & \bf 0.949 & \bf 3.04 & 0.0587 \\
\bottomrule
\end{tabular}%
}
\end{table}

\subsubsection{Effects of Dynamic Aggregation Strategy}
Next, we investigate the benefit of a time-varying strategy by comparing the static WS module with the DWS module on WavLM (Table \ref{tab:results_self-attn}). The DWS (Linguistic) consistently outperforms the static WS (Linguistic), achieving the best WER among the purely SSL feature inputs. On the other hand, the hybrid method for DWS did not show any advantage. This suggests that the module's ability to adapt to local, frame-level changes in SNR, as motivated by our MI analysis, provides a tangible performance benefit. 

\begin{figure}[t]
    \centering
    \includegraphics[width=\linewidth]{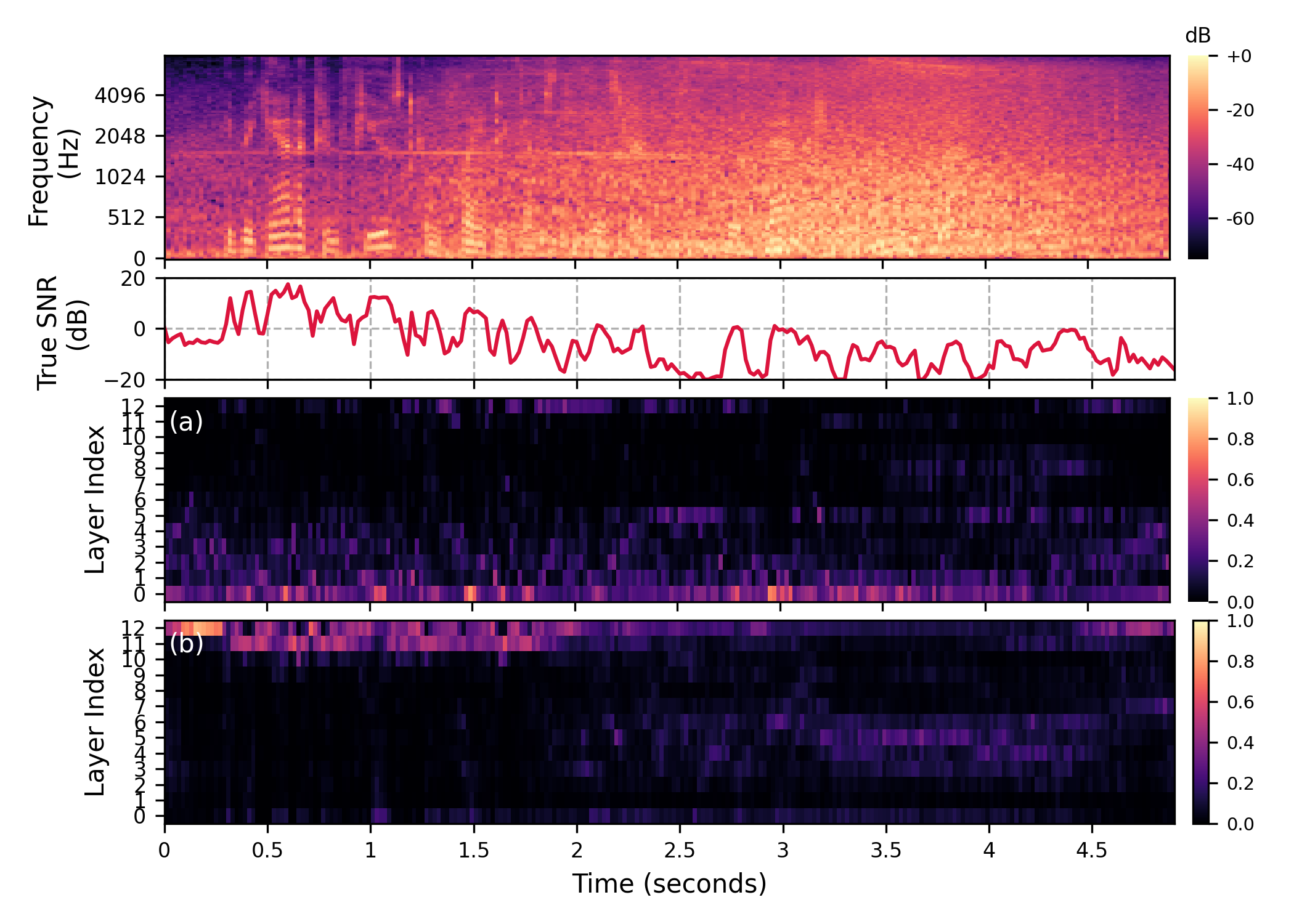}
    \vspace{-0.5cm}
       \caption{The top two subfigures show the Mel spectrogram of noisy speech and its corresponding time-varying SNR. The bottom two visualize dynamic layer weights for WavLM: (a) acoustic-tuned baseline and (b) linguistically-tuned module, respectively.}
    \label{fig:dynamic_weight_comparison}
    \vspace{-0.1mm}
\end{figure}

The visualization of the dynamic layer aggregation weights in Figure~\ref{fig:dynamic_weight_comparison} provides insight into this behavior. While the acoustically-tuned DWS (a) maintains a rigid focus on the acoustically-rich bottom layers, our Linguistically-tuned DWS (b) learns an adaptive strategy. At high SNR, the module prioritizes the top-most layers containing abstract linguistic information. Conversely, during low-SNR segments, the aggregation weights spread out to also include middle layers. 
As noise is added, the absolute magnitude of the peak is reduced (Fig. \ref{fig:mi_vs_layer_phoneme}), indicating that the top layer alone may not capture enough linguistic information. Consequently, because our linguistic module is trained to optimally combine layers, it also relies on additional layers, resulting in more distributed aggregation weights under low SNR conditions. 
A similar observation holds for the wav2vec 2 model (Fig. \ref{fig:weights_wav2vec2_col}), where the MI peak value is also low.
This suggests the module has learned a robust, time-varying policy to search for linguistic content when the primary source is degraded, validating the motivation for a dynamic approach despite the need for further refinement to maximize empirical gains.

\section{Discussion}
\label{sec:conclusion}
In this work, we analyzed the robustness of SSL speech representations using an MI framework. Our findings revealed that the distribution of the remaining phonetic information is consistently peaking in the upper layers of the model, which is similar to findings from clean-speech analysis.
And the phonetic information is degraded while noise corrupts the signal.
Based on these insights, we proposed decoupled adaptation modules that align with the original intent of using SSL features for SE, explicitly designed for linguistic roles.
The results confirm our hypothesis about a contradiction in conventional adaptation of SSL features for SE: although these features are integrated for their linguistic content, this information is corrupted by noise, and the adaptation module is trained with objectives that prioritize acoustic fidelity.
Our method instead prioritized the linguistically-rich upper layers, which were shown to be effective when the representation contains sufficient information.
As the SNR decreases, the peak information is reduced, leading the model to rely on additional layers for sufficient linguistic information and producing more distributed aggregation. Since SNR varies over time, the model needs to aggregate information locally, which supports our use of the Dynamic Weighted-Sum approach.
The combination of a decoupled framework and a dynamic architecture effectively preserves the linguistic content, which is the main purpose of using SSL representations for SE. This is achieved through a simple, well-designed aggregation of the SSL model's frozen layers, yielding significant WER improvements with minimal acoustic quality loss.

Future work could extend our MI analysis to other attributes, such as speaker identity. A further research direction is to develop alternative SE components that fully leverage the rich, abstract information contained in SSL features, moving beyond a single aggregated representation. Finally, designing a novel metric that is independent of acoustic details might be desirable to assess the semantic performance of SE.

\section{Acknowledgements}
This work was partly supported by the National Research Foundation of Korea(NRF) grant funded by the Korea government(MSIT) [No. RS-2025-24683892, 50\%], [No. RS-2024-00461617, 45\%], and Institute of Information \& communications Technology Planning \& Evaluation (IITP) grant funded by the Korea government(MSIT) [NO.RS-2021-II211343, 5\%] The GPUs were partly supported by the National IT Industry Promotion Agency (NIPA)'s high-performance computing support program in 2025.


\bibliographystyle{IEEEbib}
\bibliography{strings,refs}

\end{document}